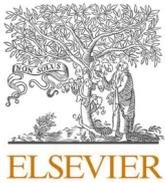
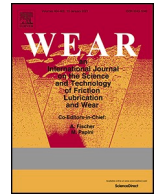

# Structural reorientation and compaction of porous MoS$_2$ coatings during wear testing

Sebastian Krauß [a,*], Armin Seynstahl [b], Stephan Tremmel [b], Bernd Meyer [c], Erik Bitzek [a,d], Mathias Göken [a], Tadahiro Yokosawa [e], Benjamin Apeleo Zubiri [e], Erdmann Spiecker [e], Benoit Merle [a]

[a] *Materials Science & Engineering I and Interdisciplinary Center for Nanostructured Films (IZNF), Friedrich-Alexander-Universität Erlangen-Nürnberg (FAU), Cauerstr. 3, 91058, Erlangen, Germany*
[b] *Engineering Design and CAD, Universität Bayreuth, Universitätsstr. 30, 95447, Bayreuth, Germany*
[c] *Interdisciplinary Center for Molecular Materials (ICMM) and Computer Chemistry Center (CCC), Friedrich-Alexander-Universität Erlangen-Nürnberg (FAU), Nägelsbachstr. 25, 91052, Erlangen, Germany*
[d] *Computational Materials Design, Max-Planck-Institut für Eisenforschung GmbH, Max-Planck-Str. 1, 40237, Düsseldorf, Germany*
[e] *Institute of Micro- and Nanostructure Research (IMN) & Center for Nanoanalysis and Electron Microscopy (CENEM), Interdisciplinary Center for Nanostructured Films (IZNF), Friedrich-Alexander-Universität Erlangen-Nürnberg (FAU), Cauerstr. 3, 91058, Erlangen, Germany*

ABSTRACT

Industrial upscaling frequently results in a different coating microstructure than the laboratory prototypes presented in the literature. Here, we investigate the wear behavior of physical vapor deposited (PVD) MoS$_2$ coatings: A dense, nanocrystalline MoS$_2$ coating, and a porous, prismatic-textured MoS$_2$ coating. Transmission electron microscopy (TEM) investigations before and after wear testing evidence a crystallographic reorientation towards a basal texture in both samples. A basal texture is usually desirable due to its low-friction properties. This favorable reorientation is associated to a tribological compaction of the porous specimens. Following running-in, sliding under high contact pressure ultimately leads to a wear rate as small as for an ideal grown bulk MoS$_2$ single crystal grown by chemical vapor deposition (CVD). This suggests that the imperfections of industrial grade MoS$_2$ coatings can be remediated by a suitable pretreatment.

## 1. Introduction

MoS$_2$ solid lubrication coatings are commonly used for providing dry lubrication in bearings under challenging environmental conditions, e. g., vacuum, high load, and irradiation [1–4]. Their lubrication performance highly depends on the microstructure [5–7] that results from the deposition process. Depending on the process parameters, sputter deposition mostly yields three types of microstructures: (i) A basal texture, where the basal planes of the hexagonal 2H–MoS$_2$ lattice [8–10] are oriented parallel to the substrate, which is believed to provide the most effective lubrication but is rarely found [11–13]. (ii) A randomly oriented, nanocrystalline microstructure, resulting in fairly good lubrication [13,14]. (iii) A prismatic texture associated to the growth of individual MoS$_2$ sheets in the vertical direction, which is generally thought to result in poorer lubrication [15–19]. Due to its dendritic growth [14], the latter includes interdendritic voids. The microstructural difference between both kinds of coatings is therefore twofold: porosity and texture. Obtaining the preferred basal texture and a dense coating requires a tight control of the deposition parameters [20]. This is very feasible under laboratory conditions but mostly beyond reach for industrial-scale processes associated with dynamic conditions [21]. Large-scale MoS$_2$ coatings therefore generally show some degree of porosity. These porous coatings were sometimes reported to undergo local compaction associated to a certain extent of crystallographic reorientation under tribological loading [13,22–24]. This gives some hope that these coatings can provide satisfactory lubrication following running-in.

In this paper we examine the compaction and reorientation of a vertically oriented, porous coating after short tribological testing compared to an inherently dense coating with nanocrystalline microstructure and to a perfectly basal oriented MoS$_2$ single crystal used as a reference. A microscale wear approach is used to determine incremental wear rates. The microstructure of the coatings is resolved via high-resolution transmission electron microscopy (HRTEM) before and after tribological testing.

## 2. Materials & methods

### 2.1. Sputter-deposited $MoS_2$ coatings and CVD single crystal

Two different magnetron sputter-deposited $MoS_2$ coatings were compared to a CVD grown $MoS_2$ single crystal. 100Cr6 bearing steel, relevant in applications, was used as substrate for the coatings. The specimens had cylindrical dimensions of Ø 30 mm × 5 mm and Ø 10 mm × 3 mm for the macroscopic and microscopic wear experiments, respectively. The substrates were tempered and hardened to 62 ± 1 HRC and fine polished with a 1 µm diamond suspension (ATM Qness GmbH, Germany) and 0.25 µm silica suspension (OP–S, Struers GmbH, Germany). Prior to placing the substrates inside the industrial-scale deposition unit (TT 300 K4; H–O-T Härte-und Oberflächentechnik GmbH & Co. KG, Germany), all substrates were ultrasonically cleaned in acetone and isopropyl alcohol for 10 min and blow dried with nitrogen. The deposition chamber was evacuated to an initial pressure of $2.5 \times 10^{-3}$ Pa before performing the PVD sputter process from hot-pressed $MoS_2$ powder targets of 99.5% purity (GfE Metalle und Materialien GmbH, Germany). Once high vacuum conditions were achieved, the substrate surface was plasma etched for 15 min. For this process a pulsed DC bias voltage (pulse frequency of 40 kHz, positive pulse duration of 5 µs) of −500 V and an argon gas flow of 500 sccm were utilized. The main difference between the deposited coatings was the rotational strategy during the process. For one batch, the substrates were positioned at a short distance of 65 mm from the target without any rotation (nanocrystalline coating), whereas for a second batch, a threefold rotation rack was operated at an average distance of 250 mm (porous coating). The first configuration mimicked laboratory conditions (laboratory grade), whereas the second configuration was closer to an industrial process (industrial grade) suitable for coating geometrically complex components. A detailed description of the deposition parameters can be found in Table 1. In order to minimize oxidation prior to testing, the samples were stored in vacuum desiccators at all times.

To compare the deposited coatings with a near perfect $2H–MoS_2$ structure, a small CVD grown single crystal (HQ graphene BV, The Netherlands) was purchased. Using the CVT (chemical vapor transport) process, the crystal was grown inside a quartz tube to dimensions of 10 mm × 8 mm and a thickness of 1 mm (single crystal).

### 2.2. Microstructural characterization

To characterize the growth mode of the specimens, the surface structure was investigated in a scanning electron microscope (SEM; Zeiss Crossbeam XB 540, Carl Zeiss AG, Germany). The investigation was carried out at an acceleration voltage of 5 kV and an ion beam current of 500 pA in secondary electron contrast. The elemental composition was furthermore investigated by energy dispersive X-ray spectroscopy (EDX) at an acceleration voltage of 20 kV and an electron current of 1.4 nA. In order to resolve the microstructure of the coatings and the bulk crystal, electron-transparent cross-section lamellae were prepared from the respective material with a focused ion beam system (FIB-SEM; Nanolab 600i, FEI Thermo Fisher Scientific Inc., USA). First, a protective platinum coating was deposited onto the surface to protect the pristine surface from gallium ion damage during milling. The protective platinum consists of a thin layer deposited with the electron beam directly on the sample surface and a thicker layer deposited with the $Ga^+$ ion beam. For coarse milling, ion beam currents ranging from 21 nA to 0.79 nA were used. After transferring the lamella onto a copper half-grid, the final thinning was performed using sequentially reduced acceleration voltages and ion beam currents down to thicknesses of ~100 nm (30 kV, 0.23 nA and 80 pA; 16 kV, 0.15 nA; 8 kV, 66 pA; 5 kV, 41 pA and 15 pA and 2 kV, 9 pA). The atomic structure of the lamellae was investigated with a double aberration-corrected Titan Themis 300 transmission electron microscope (FEI Thermo Fisher Scientific Inc., USA). Using an acceleration voltage of 300 kV, the atomic structure of the respective samples could be resolved in high-resolution transmission electron microscopy (HRTEM) mode. To investigate the local crystal structure in more detail, selected area electron diffraction (SAED) and high-angle annular dark-field scanning TEM (HAADF-STEM) techniques were used. The different polytypes of $MoS_2$ (1T, 2H and 3R) [9,25] can be differentiated by the positions/radii of the diffraction spots/rings in reciprocal space. The experimental patterns were indexed by comparisons to simulated diffraction patterns of $MoS_2$ polytypes using the software SingleCrystal (CrystalMaker Software Ltd, UK), using 2H (COD ID 9009144) and 3R (COD ID 9009148) crystal structures [26].

Surface roughness measurements were carried out using a laser scanning microscope (LSM) VK-X-150K (Keyence Corp., Japan). Nano-indentation hardness measurements were conducted in a G200 nano-indenter (KLA Inc., USA). The hardness values were averaged over 25 nanoindentations, which were carried out up to an indentation depth of 500 nm and evaluated at 5% ± 1% of the maximum indentation depth.

### 2.3. Macroscopic wear testing under vacuum conditions

The sputter-deposited coatings were tested in a rotary tribometer (HTV-SST; KTmfk, Germany) with a ball-on-disk configuration. A detailed description of the tribometer is given in Ref. [21]. The tests (n = 1) were carried out under vacuum conditions (initial pressure: 0.01 Pa) and in a rotational sliding mode with a sliding speed of 0.1 m/s. The normal force was set to 10 N, resulting in a Hertzian pressure of 1.17 GPa at the contact center. Steel balls (grade G10, ISO 3290, $R_a \leq 0.02$ µm) with a diameter of 8 mm made of bearing steel 100Cr6 (1.3505, AISI 52100) were used as counterbodies. In the tribological tests, a track radius of 11.5 mm was adjusted and the termination criterion was set at 1000 cycles to ensure that no coating failure occurred. In order to compare the friction coefficients from the macroscopic tests with those of the microscopic tests, a few tests were carried out under ambient air with the test parameters mentioned earlier. Friction coefficients were averaged over 1000 cycles. After the tribological loading, a cross-section lamella of the material under the wear track was prepared to investigate the microstructure in a deformed state. The lift-out position and process are displayed in Fig. 1. The lift-out process, as well as the TEM investigation were performed analogously to the undeformed state described in section 2.2.

### 2.4. Nanomechanical wear testing in ambient air

To compare the tribological behavior of the coatings to the ideal single crystal, nanomechanical wear experiments (n = 3) were conducted on all three samples. The tests were performed in a Nanoindenter G200 (KLA Inc., USA) equipped with a spherical diamond indenter (tip radius 50 µm, Synton MDP, Switzerland). The test sequence consisted of pressing the diamond indenter against the surface and performing a sliding motion at constant normal load. The tests were performed at normal loads of 1 mN and 10 mN, which corresponds to Hertzian contact pressures of 1.42 GPa and 3.06 GPa, respectively. The indenter was moved back and forth over a distance of 10 µm for 10000 cycles. A profiling scan at a constant load of 20 µN was performed with the indenter tip after each 1000th cycle, capturing the depth evolution of the wear track. The lateral force on the indenter tip was continuously recorded during the wear tests at 3.06 GPa. The friction coefficient was calculated as an average between cycles #9000 – #10000 over the three test runs. The total wear volume for each wear track was measured ex-situ using the LSM and the resulting wear coefficients were evaluated according to Archard [27]. Based on the indentation depth evolution and the wear volume after the test, the incremental wear volume and wear coefficient were calculated, factoring in the geometry of the diamond indenter. Additionally, the cross-section of the coatings was investigated by local FIB milling inside the wear tracks.

**Table 1**
Deposition parameters of the $MoS_2$ coatings.

| Deposition parameters | Nanocrystalline coating | Porous coating |
| --- | --- | --- |
| Rotation strategy | – | threefold |
| Target–substrate distance in mm | 65 | 250 |
| Rotational speed in rpm | – | 6 |
| Duration in s | 300 | 4200 |
| Sputtering power in kW | 1.3 | 2.0 |
| Bias voltage in V | 0 | 0 |
| Argon gas flow in sccm | 115 | 120 |
| Argon pressure in Pa | 0.70 | 0.71 |
| Chamber temperature in °C | 50 | 50 |

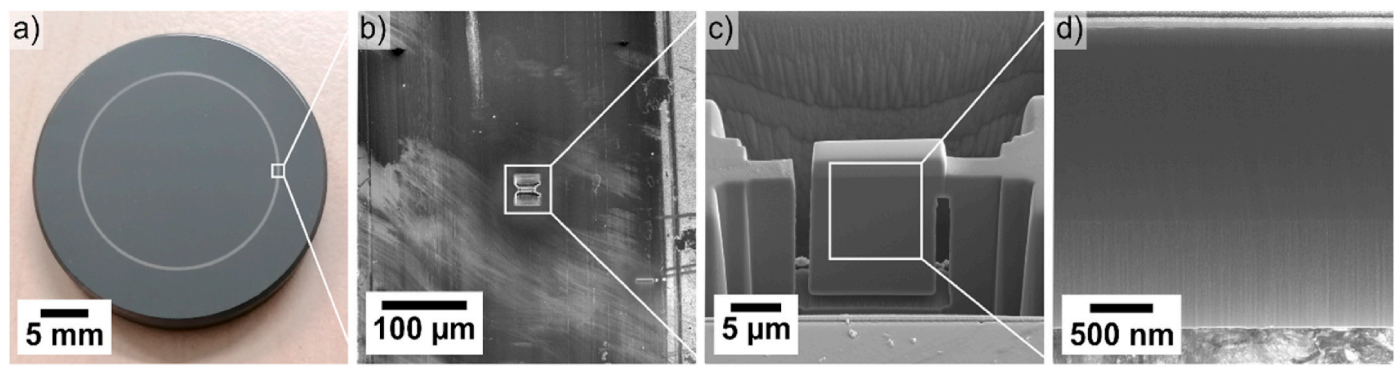

**Fig. 1.** Process and positioning of post-test lift-out. a) Sample with wear track after ball-on-disk testing. b) Lift-out position in wear track. c) Pre-cut lamella prior to lift-out. d) Final lamella in an SEM imaged in scanning transmission electron microscopy mode.

## 3. Results

### 3.1. Pristine coatings and single crystal

The properties of the pristine $MoS_2$ samples are summarized in Table 2. The elemental analysis revealed that the ratio of sulphur to molybdenum is understochiometric in all cases. A detailed explanation is given in our previous publication [21]. Laser scanning measurements of the roughness showed that all samples had very smooth surfaces ($R_a \leq 38$ nm). The indentation hardness of the nanocrystalline coating was significantly higher than for the porous coating. Please note that the hardness of the single crystal could not be quantified, because of early fracture during nanoindentation. To ease tribological comparisons between the coatings, their friction coefficients are also summarized in Table 2. They were measured as described in section 2.3 and 2.4.

### 3.2. As-deposited microstructure

The surface morphology of the PVD coatings and the $MoS_2$ single crystal are shown in Fig. 2. The surface of the nanocrystalline coating (Fig. 2a) shows a mixture of plates and filament-like structures, albeit with hardly any apparent voids. In contrast, the coating deposited with the threefold rotation (Fig. 2b) exhibits a striking open porosity. It consists of a hierarchical, dendritic surface structure separated by large voids between the dendrites. A similar surface structure of $MoS_2$ coatings has been observed in various publications, emphasizing how common porous coatings are [5,7,17,28–32]. Fig. 2c shows the smooth surface planes of the single crystal with a roughness $R_a$ of 38 nm ± 11 nm with few surface steps with a height of 330 nm ± 120 nm, marking the end of individual $MoS_2$ basal plane stacks.

The TEM investigation of lift-out lamellae provides a more detailed understanding of the microstructure of the specimens (Fig. 3). The elongated crystallites in Fig. 3a–b shows extensive lattice fringes with a lattice spacing of 6.2 Å, corresponding to the distance between two consecutive basal planes of $MoS_2$ [33]. The elongated $MoS_2$ sheets consist of the stacking of 3–10 layers. Their relative orientation to the substrate determines the coating texture. The SAED patterns of the respective coatings further assist in the assessment of the texture. Since the area probed by SAED is significantly larger than the HRTEM micrographs shown in Fig. 3, the combination of both techniques ensures reliable findings. The nanocrystalline coating (Fig. 3a) features nanometer-sized, randomly oriented crystallites of about 10 $MoS_2$ layers, similar to microstructure (ii) described in the introduction. The SAED pattern shows continuous rings, which corresponds to the expected pattern of a polycrystalline sample with randomly oriented crystallites. The porous coating (Fig. 3b) shows vertically oriented crystallites, similar to microstructure (iii). The thickness of about 6–10 layers of the $MoS_2$ sheets is comparable to the size of the nanocrystals in Fig. 3a. The length of the crystallites is determined by their respective orientation and competitive growth. In the corresponding SAED pattern, a preferred orientation is apparent as distinct peaks with higher intensity are clearly observable on top of continuous diffraction rings with lower intensity. The {0002} peaks correspond to the edge-on orientation of the $MoS_2$ sheets, which is also apparent in the HRTEM image in Fig. 3b. In contrast, the diffraction spots on the $\{01\bar{1}0\}$ diffraction ring stem from similar crystallites, albeit projected along the [0001] zone axis. Both observations are in agreement with a prismatic texture of the coating.

**Table 2**
Properties of the pristine $MoS_2$ samples.

| Parameter | Nanocrystalline coating | Porous coating | Single crystal |
| --- | --- | --- | --- |
| S/Mo Ratio | 1.59 | 1.67 | 1.8 |
| Roughness Ra in nm | 29 ± 6 | 25 ± 4 | 38 ± 11 |
| Indentation hardness in GPa | 5.69 ± 0.54 | 0.09 ± 0.03 | – |
| Friction coefficient | | | |
| macroscopic, vacuum | 0.029 ± 0.003 | 0.035 ± 0.005 | – |
| macroscopic, ambient air | 0.08 ± 0.01 | 0.06 ± 0.01 | – |
| microscopic, ambient air | 0.12 ± 0.01 | 0.16 ± 0.01 | 0.04 ± 0.01 |

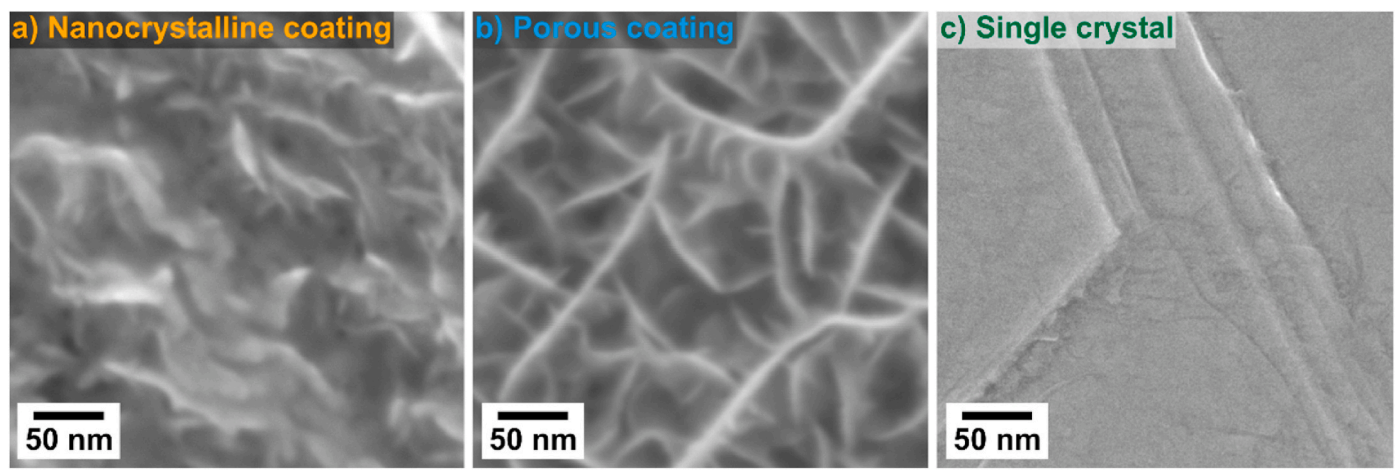

**Fig. 2.** Pristine surface structure of the samples in secondary electron (SE) contrast. a) Nanocrystalline coating. b) Porous coating. c) Bulk single crystal. The samples show strikingly different surface structures: A dense, plate-like surface for the nanocrystalline coating, dendritic crystallites with open porosity for the porous coating and a flat surface with surface steps for the single crystal.

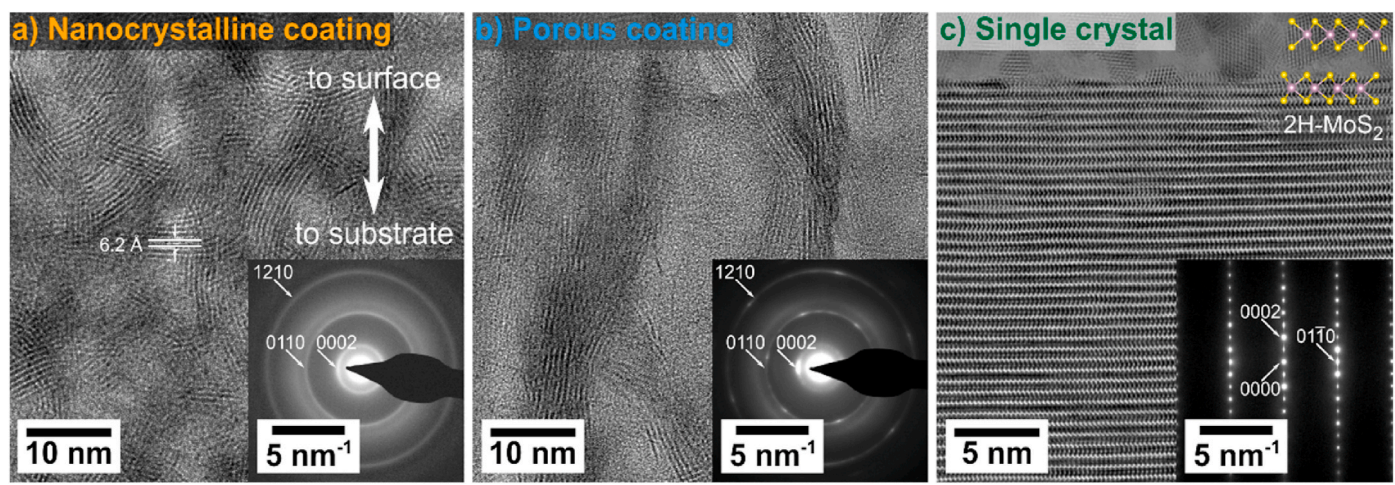

**Fig. 3.** Nanomorphology and crystalline structure of $MoS_2$ films studied by high-resolution transmission electron microscopy and selected area electron diffraction (insets). a) Nanocrystalline coating. b) Porous coating. c) CVD-grown $2H–MoS_2$ single crystal investigated along the $[\bar{2}110]$ zone axis with included crystal structure model. The structure of the coatings differs strongly from a perfect single crystal.

Depending on the local orientation of the crystallites in the plane of the film, the {0002} basal planes appear in edge-on or in-plane orientation. The 34.4° angular range of the (0002) peak in the SAED pattern indicates that the crystallites are oriented between ± 17.2° from the surface normal. Fig. 3c shows the cross-section of the CVD single crystal oriented along $[\bar{2}110]$ zone axis (see SAED pattern inset). The $MoS_2$ planes are well aligned, maintaining perfect $2H–MoS_2$ geometry except for the presence of few dislocations, which is in good agreement with microstructure (i) from the introduction.

Indexing of the experimentally acquired SAED patterns in Fig. 3 is performed by a comparison to the expected positions of the diffraction spots of 2H–MoS$_2$ (using database COD ID 9009144) [26] provided in Table 3. A more detailed indexing of the SAED patterns is given in the supplementary information. The radii of the two innermost rings in Fig. 3a–b match the reciprocal lattice spacing for (0002) and (01$\bar{1}$0). (0002) being the basal plane of MoS$_2$, the aforementioned preferred orientation of the porous coating corresponds to a prismatic texture. The texture change after tribological testing is addressed in the next section.

**Table 3**
Experimentally determined reciprocal lattice spacings from the SAED patterns and expected values (COD ID 9009144) given in Å$^{-1}$.

| Lattice plane | Nanocrystalline coating | Porous coating | Single crystal | Database "COD ID 9009144" |
| --- | --- | --- | --- | --- |
| {0002} | 0.159 ± 0.001 | 0.160 ± 0.001 | 0.162 ± 0.001 | 0.1627 |
| $\{01\bar{1}0\}$ | 0.379 ± 0.002 | 0.382 ± 0.001 | 0.371 ± 0.001 | 0.3654 |
| After deformation | | | | |
| {0002} | 0.157 ± 0.001 | 0.166 ± 0.002 | – | 0.1627 |
| $\{01\bar{1}0\}$ | 0.375 ± 0.001 | 0.379 ± 0.001 | – | 0.3654 |

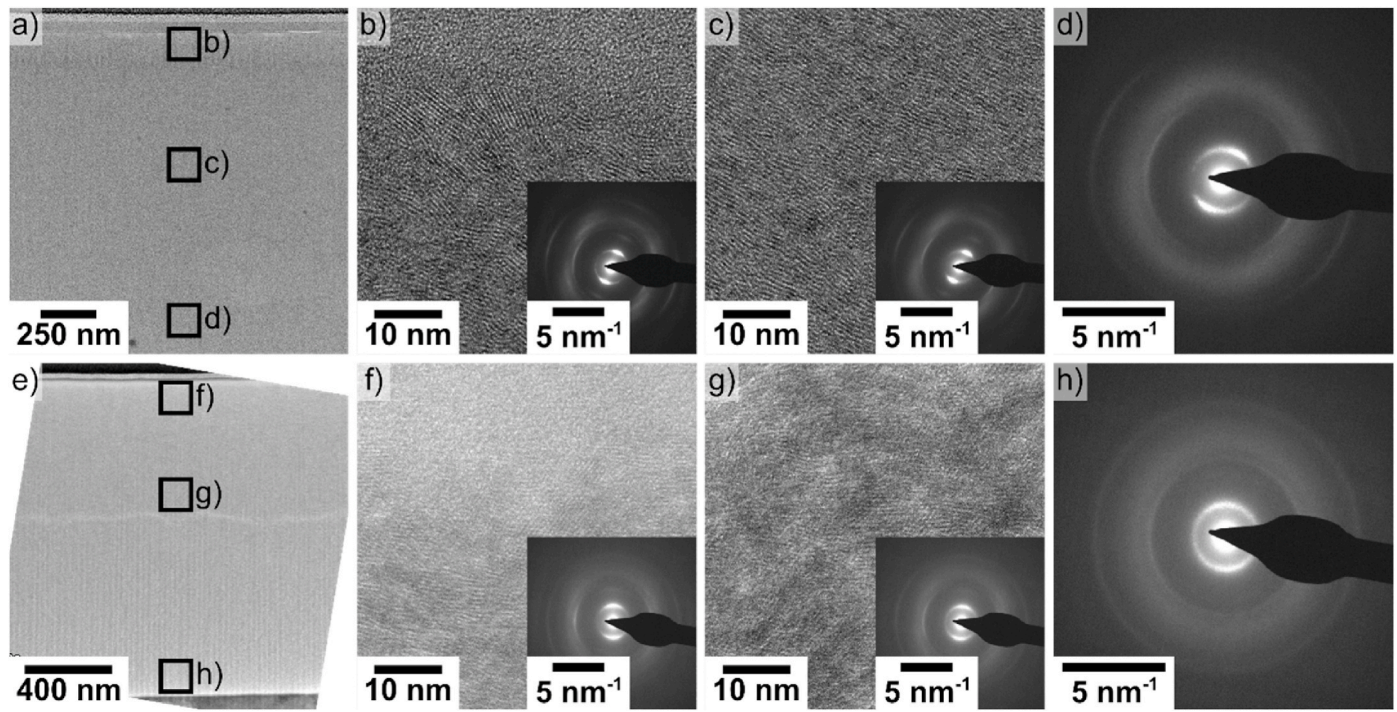

**Fig. 4.** Microstructural HRTEM and SAED investigations after short wear testing. a-d) Nanocrystalline coating. e-h) Porous coating. After tribological testing, the nanocrystalline and porous coatings are reoriented into a predominantly basal texture.

### 3.3. Microstructure after macroscopic wear testing in vacuum

The microstructural TEM characterization after short wear testing of the nanocrystalline coatings is shown in Fig. 4a–d. The mechanical loading resulted in the formation of a texture. Compared to Fig. 3a, the basal planes (see Fig. 4b) show a more uniform orientation, which is roughly parallel to the surface. The small inclination arises from the off-center position of the FIB lift-out within the curved wear track. The reorientation extends deep into the coating, as evidenced by Fig. 4c. This conclusion is also supported by the SAED inset, where the intensities in the first and second ring, corresponding to (0002) and (01$\bar{1}$0), have clearly changed to a stronger texture with highly localized ring fragments (compare to Fig. 3). The texture can still be recognized close to the substrate in Fig. 4d, but is less pronounced than near the surface. This effect is evidenced by the increase in angular range of the (0002) ring fragment in the SAED patterns, from ±19.3° at subsurface to ± 20.4° in the center of the coating and to ± 40.4° close to the substrate. This shows that the preferred orientation of the MoS$_2$ sheets is progressively weakening in deeper coating areas.

An overview TEM micrograph of the initially porous coating in worn state is shown in Fig. 4e, where the positions of the HRTEM and SAED measurements of Fig. 4f–h are indicated. The horizontal artefact, as well as the vertical curtaining, originate from FIB milling and should be disregarded. Similar to the polycrystalline coating, Fig. 4f as well as Fig. 4g show significant reorientation after short wear testing. $MoS_2$ crystallites in vertical orientation cannot be found anymore in the deformed state. Instead, the microstructure is rearranged to a basal texture with strong character in the upper part of the lamella and with a weaker character 600 nm below the coating surface. Near the substrate, a preferred orientation is almost completely lacking (Fig. 4h). The angular range of ±21.7°, 25.28° and 33.43° for the subsurface, center and substrate-near areas evidence an increasing misorientation deeper into the coating, analogous to the nanocrystalline coating. Note that due to the larger thickness of the TEM lamella in the deeper investigated area, the electron beam travels through a thicker layer, which leads to a lesser crystallinity in Fig. 4g–h, due to more diffuse and inelastic scattering intensities in the HRTEM images and SAED patterns.

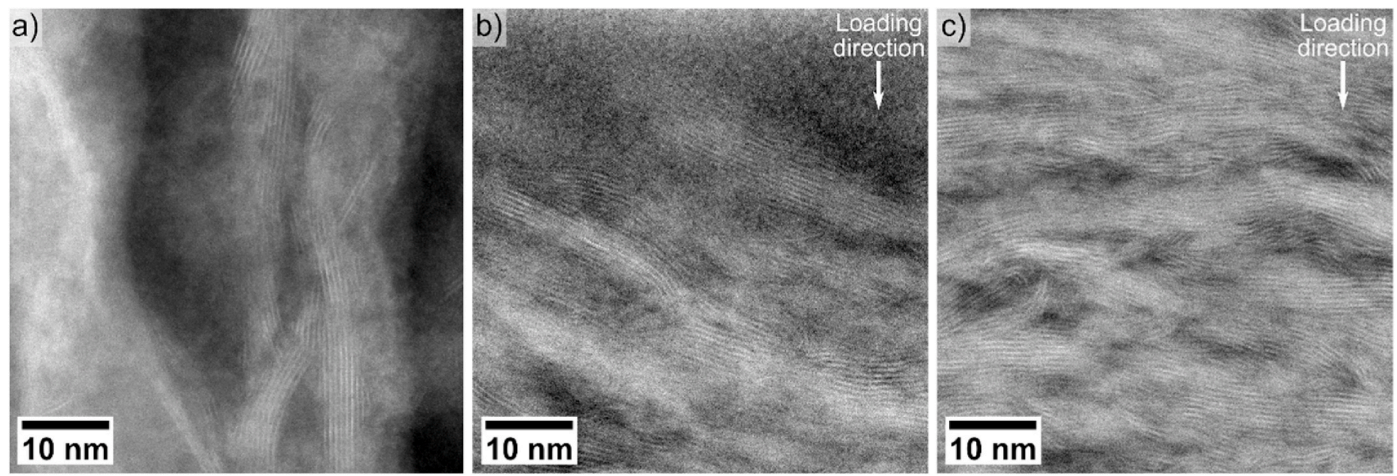

**Fig. 5.** Porosity evolution in the porous coating in HAADF-STEM contrast. a) Initial state prior to tribological loading. b) Upper part of the lamella after short tribological loading. c) Middle part of the lamella after short tribological loading. The porous coating shows significant compaction after tribological testing.

The residual porosity after tribological loading in the vertically textured coating is evaluated from Fig. 5. The HAADF-STEM contrast shows high intensity where atoms with high atomic number (Z contrast) are located and low intensity where either atoms with lower atomic number or fewer atoms are located. Therefore, the intensity distribution in HAADF-STEM imaging is thickness dependent and the contrast can be used to identify pores. The pores with an approximately ellipsoidal shape (Fig. 5a) have a size of 12 nm × 50 nm in the initial state of the coating, which is effectively reduced after short tribological loading to 2 nm × 5 nm. This is the case for both the surface-near area (Fig. 5b) and below (Fig. 5c). The vertical crystallites are reoriented, stacked and compacted by the wear process. Additionally, the pores are reshaped to elongated horizontal voids in-between compacted crystallites.

For both coatings, an approximately 20 nm thick amorphous surface layer is identified, which could be a result of multiple effects, such as surface damage from wear testing, the reincorporation of wear debris, surface damage from FIB milling or a combination of these.

### 3.4. Nanomechanical wear testing at different contact pressures in ambient air

The results from the nanomechanical sliding wear tests are depicted in Fig. 6 for increasing Hertzian contact pressures: 1.42 GPa (Fig. 6a) and 3.06 GPa (Fig. 6b). The tests on the single crystalline sample show no significant wear rate at all over the whole course of the experiment at 1.42 GPa. By increasing the pressure to 3.06 GPa, this trend changes. There is a significant amount of volume loss during the first 1000 cycles. During the next 1000 cycles, the worn volume decreases dramatically and remains negligible after 3000 cycles. The porous and nanocrystalline samples follow similar global trends at a pressure of 3.06 GPa, with an initial transient associated with a high wear coefficient and a follow-up steady-state phase associated with lower values. This initial transient is visible for the porous coating at a Hertzian pressure of 1.42 GPa, but the nanocrystalline coating hardly shows any initial transient. A pronounced initial transient is identified on the nanocrystalline coating at 3.06 GPa. The wear ultimately becomes negligible after 3000 to 4000 cycles. The porous coating shows an overall higher wear coefficient at a contact pressure of 1.42 GPa, especially under a steady-state phase lasting until the interruption of the test. At 3.06 GPa, the steady-state sets in as fast as the nanocrystalline coating and the wear coefficient decreases to values as low as with the single crystal. Some datapoints represent negative wear rates, which means that volume is introduced into the system. This artefact can be traced back to wear debris being pushed into the track during profiling. One representative residual wear track on each sample is shown in Fig. 7. The wear tracks in the figures should be compared to the cumulative wear volume measured throughout the test.

The nanocrystalline coating is associated to the strongest formation of wear debris (Fig. 7a) under a Hertzian contact pressure of 1.42 GPa. The dark contrast in the image corresponds to surface contamination or oxidation, since the coating was exposed to ambient air for several days during and after wear testing. The track on the porous sample (Fig. 7b) is much wider than those of the single crystal and the nanocrystalline coating. The wear track on the single crystal (Fig. 7c) shows a very smooth surface with little deformation and no significant wear debris.

By increasing the contact pressure to 3.06 GPa on the nanocrystalline coating, a large amount of wear debris formed, which are separated from the wear track by up to 20 µm (Fig. 7d). Furthermore, in the center of the track an inhomogeneity of the tribological film is visible, where only a thin lubrication layer remains on the substrate. The porous coating shows a larger track width, resulting from a deeper indentation (Fig. 7e). The single crystal shows a deformed wear track with pile-ups at both sliding direction ends (Fig. 7f). Overall, the largest volume of coating is lost during wear on the porous coating, despite the fact that only a low amount of wear debris has formed, hinting at a different abrasion behavior than for the other samples.

Cross-sections of the wear tracks are provided in Fig. 8. As the wear tracks are several micrometers wide, only the center of the track is shown in the cross-sectional views. For the nanocrystalline sample (Fig. 8a), the surface of the wear track appears to be very smooth and a clear residual indent is observable for a contact pressure of 1.42 GPa. The porous coating shows a strongly densified surface-near region (Fig. 8b,e). This area extends further through the coating as the contact pressure is increased. The surface of the wear track is rougher compared to the tests on the single crystal. For the single crystal no significant abrasion is observable for testing at 1.42 GPa (Fig. 8c) and a small residual indent is present at a Hertzian pressure of 3.06 GPa (Fig. 8f). As the contact pressure increases (Fig. 8d), the nanocrystalline coating is almost entirely removed, with only a very thin residual coating remaining at the substrate interface. Note that the indentation depth after testing at 3.06 GPa is about the same ($\approx$1 µm) for the nanocrystalline and the porous coating, while the initial coating thickness for the porous coating (2.4 µm) was higher than for the nanocrystalline coating (1.2 µm).

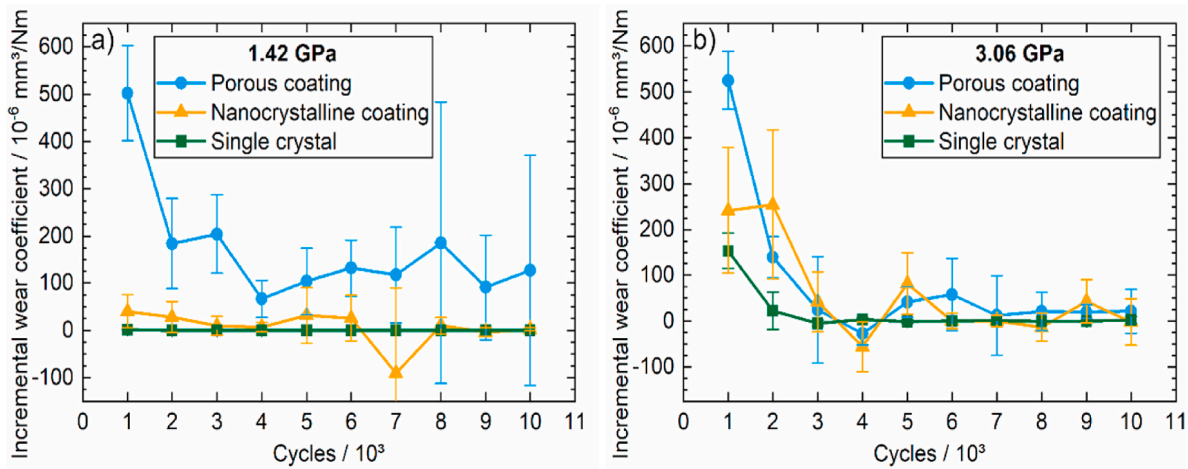

**Fig. 6.** Incremental wear coefficient from nanomechanical wear testing in ambient air. a) Testing at a Hertzian contact pressure of 1.42 GPa. b) Testing at a Hertzian contact pressure of 3.06 GPa. After initial transition, the samples reach a steady-state.

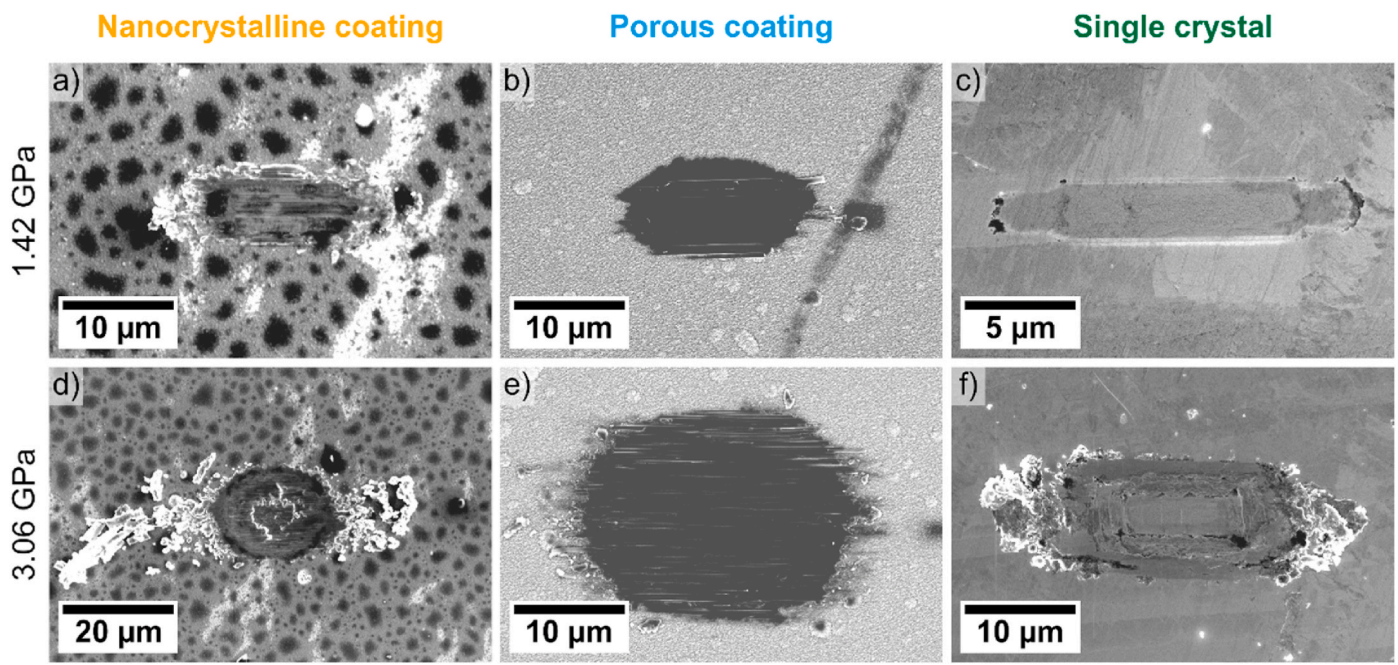

**Fig. 7.** Representative residual wear tracks after nanomechanical wear testing. a-c) Tests at Hertzian contact pressures of 1.42 GPa. d-f) Tests at Hertzian contact pressures of 3.06 GPa. The width of the residual wear track is a direct measure for the extruded volume, since the width is determined by the indentation of the spherical counterbody. Note the different scale in c) and d).

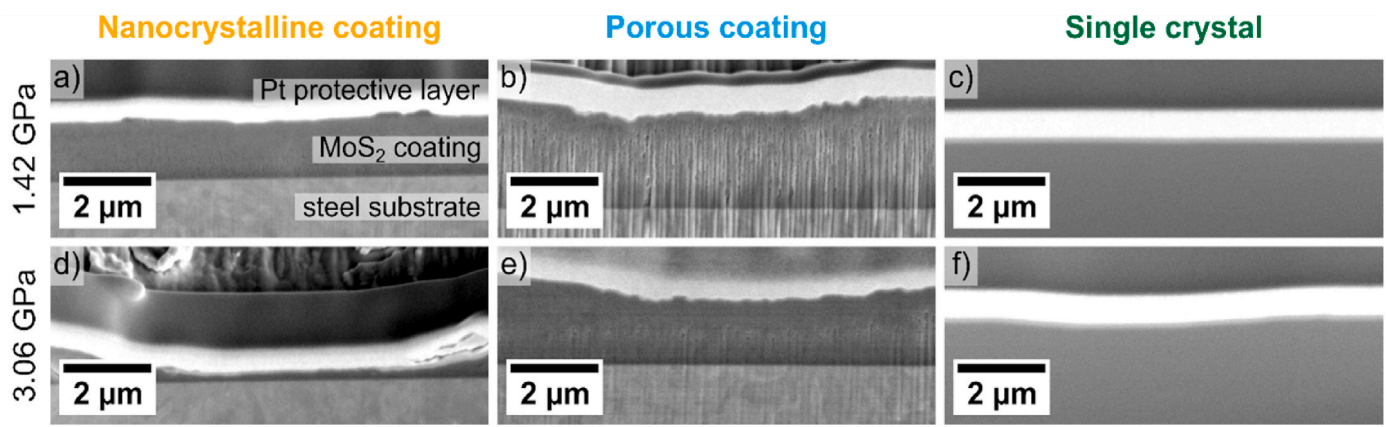

**Fig. 8.** FIB milled cross-sections of the residual coating thickness below the wear tracks. View direction is parallel to the sliding direction. a-c) Tests at Hertzian contact pressures of 1.42 GPa. d-f) Tests at Hertzian contact pressures of 3.06 GPa. The nanocrystalline coating shows a dense and homogenous lubrication layer, while the porous coating shows contact pressure dependent compaction.

## 4. Discussion

### 4.1. Porosity

The coating deposited under industrial conditions is characterized by its porosity. In the initial state, pores are present as vertical, interdendritic voids. They result from the dendritic growth of $MoS_2$, which itself originates from: (i) faster growth of $MoS_2$ crystallites within their basal (0001) plane, compared to the out-of-plane direction [0001] [31], and (ii) branching of the crystallites, due to unobstructed growth in non-horizontal directions [14]. This branching behavior is not found in our nanocrystalline coatings because of the shorter target-to-substrate distance during deposition: due to their higher kinetic energy, the incoming atoms would destroy any protruding dendrite in formation when hitting the sample surface [34,35]. At lower kinetic energy, the protruding dendrites lead to shadowing during sputter coating and result in voids being embedded within the dendrite cores.

### 4.2. Compaction

From the pre- and post-wear test TEM imaging, a compaction effect is clearly evidenced for the porous coating. Little $MoS_2$ material is lost during this process, as evidenced by the near absence of debris in Fig. 7b, e. This means that the original amount of solid lubricant is preserved for follow-up operations. As will be shown below, this compaction plays a major role in the tribological behavior of these films. It is also worth mentioning that the filling process of the pores is not yet fully completed after 1000 cycles (ball-on-disk testing at 1.17 GPa), since small residual pores are still visible in the cross-sections, see Fig. 5.

Nanocrystalline films deposited under more ideal laboratory conditions are not affected by the compaction, because they are dense from the onset.

*4.3. Structural reorientation*

Both investigated $MoS_2$ coatings show a significant reorientation after short wear testing. The initial random orientation of the nanocrystalline coating is turned into a basal texture during tribological loading. Similar reorientation under tribological loading has been observed in Refs. [13,22,23,36–39]. The process is assisted by the easy glide of $MoS_2$ sheets with respect to each other. This reorientation corresponds to a self-optimization of the coating, as the basal texture corresponds to the lowest energy configuration [40] (fewer random high angle grain boundaries). A similar process is at play in the porous coating, albeit to a lesser extent. Here, it is likely that the initially vertical $MoS_2$ crystallites break apart under the applied stress and that the resulting products are stacked together to fill the existing voids. During this process, the energy minimization driving force is in competition with the existing geometry of the voids to be filled. This likely explains why the reorientation yields an imperfect basal texture, see Figs. 5 and 4b,f.

Compared to literature, the experiments in this research are very short. A previous study on a similar porous coating showed a decrease in friction over the first 14300 cycles [21] during ball-on-disk tests, hinting that the reorientation process might not yet be fully completed here.

The reorientation appears to initiate from the surface and extend progressively towards the substrate. It cannot be a consequence of the Hertzian stress, which is roughly constant throughout the coating thickness while the shear stress maximum is located well below (ca. 30 μm for steel on steel contact with $\nu = 0.3$ [41]). This points at contact friction being the likeliest trigger of the reorientation process.

*4.4. Apparent wear rate*

The porous coating experiences a higher apparent wear rate than its dense counterparts at a low contact pressure, see Fig. 6a. This is likely a result of the compaction process. The indentation of the counterbody is not solely due to material removal during the wear process, but also to the collapse of pores beneath it, leading to the compaction of the coating. Its contribution to apparent wear can be assessed by comparing the cross-sections and residual depths of the porous (178 nm ± 8 nm) and fully dense nanocrystalline (45 nm ± 15 nm) coatings after 1000 cycles under a contact pressure of 1.42 GPa. Assuming that abrasion takes place to the same extent, the difference of 132 nm, i.e 74 %, should be attributed to compaction.

The higher apparent wear value is only found at low pressures. This is likely due to the contact-pressure dependent progress of compaction. As seen in Fig. 8, the compacted zone does not extend through the whole coating thickness until the high-stress regime. When this is the case, we presume that more material is transformed into a low-friction configuration, which could explain why the corresponding wear rates approach the performance of dense coatings, both here and during macroscopic testing [21].

*4.5. Influence of pile-up behavior*

The magnitude of the abrasion behavior can be assessed from the pile-up formed at the edges of the wear tracks, see Fig. 7. While the single crystalline sample and the nanocrystalline coating form pile-ups, the porous coating allows the counterbody to sink into the coating surface. Intrinsically, pile-ups increase the contact area between the lubricant and the counterbody, leading to higher friction, which is detrimental to the wear performance. However, the sink-in behavior of the porous coating allows the counterbody to penetrate much deeper into the coating, which leads to a higher contact area as well, but effectively hinders wear debris formation. Note that the pile-up formation at the end of the wear tracks caused by the repetitive linear motion of the counterbody would not occur under circular motion, e.g. ball-on-disk testing. There, debris could be reincorporated (with the exception of lateral pile-ups). Therefore, pile-up formation in dense coatings is not critically detrimental to the wear behavior but still remains a high-friction configuration.

*4.6. Influence of oxygen*

Since the microscale wear experiments were performed in ambient air, a certain oxidation contribution to the wear behavior is expected for all samples. The presence of oxygen is usually believed to increase friction and lower the lubrication performance of $MoS_2$ solid lubricant coatings due to $MoO_3$ formation [38,42]. The oxidation is expected to be more pronounced and hinders easy shear when multiple basal edge sites are exposed [6,43]. From the active sites model [6,44], the bulk single crystal investigated here is hardly expected to suffer from oxidation because of its basal surface orientation. For the nanocrystalline coating, the random crystal orientation offers only few oxygen adsorption sites, which are effectively transformed to a less reactive state by the reorientation process. For the porous coating, the expectation is that a large reactive surface area is exposed to the ambient air because of the open porosity. Regardless, the effect of oxidation is obvious for the frictional behavior, but is likely to play only a minor role in the microstructural evolution, since porosity is closed via compaction as soon as the test is started. The follow-up reorientation of the crystal structure to a basal texture protects the coating from further oxidation.

*4.7. Deformation models*

Our observations can be summarized with the sketches shown in Fig. 9.

For an inherently dense coating, the material removal processes are mostly limited to abrasion and plastic deformation, producing large amounts of wear debris, forming pile-ups and compacting them at either end of the sliding track. Most of the accumulating deformation takes place during an initial transient, see Fig. 6. Furthermore, a reorientation of the underlying lubrication layer sets in, which ultimately results in a lower wear rate during steady-state deformation.

For porous coatings, the transition is dominated by compaction and less material is removed, which leads to lower wear debris formation and an effective sink-in of the counterbody. We believe that after transition, the combination of reduced wear rate and the reincorporation of wear debris particles leads to a steady state characterized by a low wear rate. Therefore, porous coatings may be suitable for tribological applications, provided a proper running-in process is executed. The coatings could for instance be deposited to a larger thickness and pre-compacted prior to the application. In this way, porous coatings deposited under industrial conditions could achieve a lifetime equivalent to denser prototype specimens.

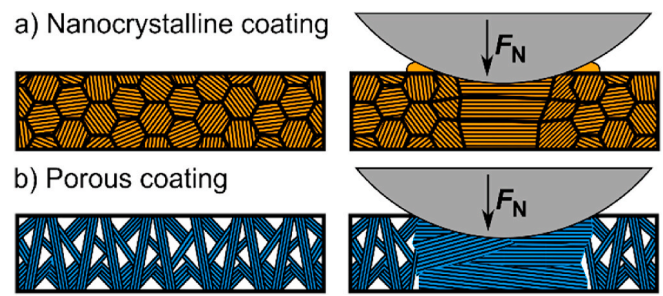

**Fig. 9.** Schematic representation of the compaction and reorientation behavior of a) inherently dense and b) porous, vertically oriented coatings after short tribological testing. Nanocrystalline coatings are worn by abrasive processes, where porous coatings show significant compaction. Both types of coatings reorient into a low-friction basal-textured state.

## 5. Conclusions

The following conclusions can be drawn from the investigations:

 I. The wear behavior of sputter-deposited $MoS_2$ coatings can be divided into an initial transition (running-in stage) showing a high wear rate, structural reorientation and compaction and a steady state stage with constant, low wear rate.
 II. During transition, prismatic-textured $MoS_2$ coatings undergo a structural reorientation into a basal-textured, low-friction configuration.
 III. High porosity coatings are significantly compacted during transition and show a comparable wear rate as inherently dense coatings after completed compaction and reorientation.
 IV. To achieve low wear rates on porous coatings, a high load is necessary to reach a high stress regime, where a low-friction orientation is predominant.
 V. Nanocrystalline as well as porous $MoS_2$ coatings, resulting from laboratory-scale and industrial-scale deposition, show a rapid decay of the apparent wear rate down to values comparable with a basal-oriented single crystalline $MoS_2$ sample.
 VI. The high wear volume during transition can be attributed to material abrasion from the surface for coatings with low porosity and to compaction processes for high porosity coatings.
 VII. Porous coatings allow the counterbody to sink into the coating, preventing wear debris formation and resulting in increase in friction.

## Credit author statement

Sebastian Krauβ: Methodology, Validation, Formal analysis, Investigation, Data curation, Writing – original draft, Writing – review & editing. Armin Seynstahl: Methodology, Investigation, Writing – original draft, Writing – review & editing. Stephan Tremmel: Methodology, Resources, Conceptualization, Writing – review & editing, Supervision, Funding acquisition, Project administration. Bernd Meyer: Conceptualization, Writing – review & editing, Funding acquisition. Erik Bitzek: Conceptualization, Writing – review & editing, Funding acquisition. Mathias Göken: Writing – review & editing. Tadahiro Yokosawa: Investigation, Data curation, Writing – review & editing. Benjamin Apeleo Zubiri: Methodology, Validation, Writing – review & editing. Erdmann Spiecker: Resources, Writing – review & editing. Benoit Merle: Conceptualization, Methodology, Formal analysis, Writing – review & editing, Supervision, Funding acquisition, Project administration.

## Declaration of competing interest

The authors declare that they have no known competing financial interests or personal relationships that could have appeared to influence the work reported in this paper.

## Acknowledgements

This research was funded by the German Research Foundation (DFG) Priority Program SPP 2074 "Fluid-free lubrication systems with high mechanical loads", grant number (GEPRIS) 407707942 (ME 4368/7-1, ME 2670/8-1 and TR 1043/7-1). S.K. and E.B. acknowledge partial funding from the European Research Council (ERC) under the European Union's Horizon 2020 research and innovation program (microKIc-Microscopic Origins of Fracture Toughness, grant agreement no. 725483). A. S. and S. T. thank S. Wartzack from Engineering Design (FAU) for the opportunity to use resources. This research used resources from the Center for Nanoanalysis and Electron Microscopy (CENEM) at Friedrich-Alexander-Universität Erlangen-Nürnberg (FAU).

## Appendix A. Supplementary data

Supplementary data to this article can be found online at https://doi.org/10.1016/j.wear.2022.204339.

## References

[1] F. Koch, R. Nocentini, B. Heinemann, S. Lindig, P. Junghanns, H. Bolt, MoS2 coatings for the narrow support elements of the W-7X nonplanar coils, Fusion Eng. Des. 82 (2007) 1614–1620, https://doi.org/10.1016/j.fusengdes.2007.07.027.
[2] M.R. Vazirisereshk, A. Martini, D.A. Strubbe, M.Z. Baykara, Solid lubrication with MoS2: a review, Lubricants 7 (2019) 57, https://doi.org/10.3390/lubricants7070057.
[3] O. Krause, F. Müller, S. Birkmann, A. Böhm, M. Ebert, U. Grözinger, T. Henning, R. Hofferbert, A. Huber, D. Lemke, R.-R. Rohloff, S. Scheithauer, T. Gross, T. Fischer, G. Luichtel, H. Merkle, M. Übele, H.-U. Wieland, J. Amiaux, R. Jager, A. Glauser, P. Parr-Burman, J. Sykes, High-precision Cryogenic Wheel Mechanisms of the JWST/MIRI Instrument: Performance of the Flight Models, California, USA, San Diego, 2010, p. 773918, https://doi.org/10.1117/12.856887.
[4] G. Tsuyuki, R. Reeve, Galileo high-gain antenna deployment anomaly thermal analysis support, J. Thermophys. Heat Tran. 9 (1995) 771–777, https://doi.org/10.2514/3.737.
[5] V. Buck, Preparation and properties of different types of sputtered MoS2 films, Wear 114 (1987) 263–274, https://doi.org/10.1016/0043-1648(87)90116-5.
[6] P.D. Fleischauer, Effects of crystallite orientation on environmental stability and lubrication properties of sputtered MoS2 thin films, OR Trans. 27 (1984) 82–88, https://doi.org/10.1080/05698198408981548.
[7] D. Kokalj, J. Debus, D. Stangier, H. Moldenhauer, A. Nikolov, A. Wittig, A. Brümmer, W. Tillmann, Controlling the structural, mechanical and frictional properties of MoSx coatings by high-power impulse magnetron sputtering, Coatings 10 (2020) 755, https://doi.org/10.3390/coatings10080755.
[8] B.J. Mrstik, R. Kaplan, T.L. Reinecke, M. Van Hove, S.Y. Tong, Surface-structure determination of the layered compounds MoS2 and NbSe2 by low-energy electron diffraction, Phys. Rev. B 15 (1977) 897–900, https://doi.org/10.1103/PhysRevB.15.897.
[9] M.A. Macchione, R. Mendoza-Cruz, L. Bazán-Diaz, J.J. Velázquez-Salazar, U. Santiago, M.J. Arellano-Jiménez, J.F. Perez, M. José-Yacamán, J.E. Samaniego-Benitez, Electron microscopy study of the carbon-induced 2H–3R–1T phase transition of MoS2, New J. Chem. 44 (2020) 1190–1193, https://doi.org/10.1039/C9NJ03850G.
[10] Z. Dai, W. Jin, M. Grady, J.T. Sadowski, J.I. Dadap, R.M. Osgood, K. Pohl, Surface structure of bulk 2H-MoS2 (0001) and exfoliated suspended monolayer MoS2: a selected area low energy electron diffraction study, Surf. Sci. 660 (2017) 16–21, https://doi.org/10.1016/j.susc.2017.02.005.
[11] X. Zhang, R. Vitchev, W. Lauwerens, L. Stals, J. He, J.-P. Celis, Effect of crystallographic orientation on fretting wear behaviour of MoSx coatings in dry and humid air, Thin Solid Films 396 (2001) 69–77, https://doi.org/10.1016/S0040-6090(01)01141-5.
[12] C. Müller, C. Menoud, M. Maillat, H.E. Hintermann, Thick compact MoS2 coatings, Surf. Coating. Technol. 36 (1988) 351–359, https://doi.org/10.1016/0257-8972(88)90165-X.
[13] B. Vierneusel, T. Schneider, S. Tremmel, S. Wartzack, T. Gradt, Humidity resistant MoS2 coatings deposited by unbalanced magnetron sputtering, Surf. Coating. Technol. 235 (2013) 97–107, https://doi.org/10.1016/j.surfcoat.2013.07.019.
[14] J. Moser, F. Lévy, Growth mechanisms and near-interface structure in relation to orientation of MoS2 sputtered thin films, J. Mater. Res. 7 (1992) 734–740, https://doi.org/10.1557/JMR.1992.0734.
[15] C. Muratore, A.A. Voevodin, Control of molybdenum disulfide basal plane orientation during coating growth in pulsed magnetron sputtering discharges, Thin Solid Films 517 (2009) 5605–5610, https://doi.org/10.1016/j.tsf.2009.01.190.
[16] P.A. Bertrand, Orientation of rf-sputter-deposited MoS2 films, J. Mater. Res. 4 (1989) 180–184, https://doi.org/10.1557/JMR.1989.0180.
[17] B. Vierneusel, S. Tremmel, S. Wartzack, Effects of deposition parameters on hardness and lubrication properties of thin MoS2 films, Mater. Werkst. 43 (2012) 1029–1035, https://doi.org/10.1002/mawe.201200942.
[18] J. Moser, H. Liao, F. Levy, Texture characterisation of sputtered MoS2 thin films by cross-sectional TEM analysis, J. Phys. Appl. Phys. 23 (1990) 624–626, https://doi.org/10.1088/0022-3727/23/5/026.
[19] M.R. Hilton, P.D. Fleischauer, TEM lattice imaging of the nanostructure of early-growth sputter-deposited MoS2 solid lubricant films, J. Mater. Res. 5 (1990) 406–421, https://doi.org/10.1557/JMR.1990.0406.
[20] J. Wang, W. Lauwerens, E. Wieers, L.M. Stals, J. He, J.P. Celis, Effect of power mode, target type and liquid nitrogen trap on the structure and tribological properties of MoSx coatings, Surf. Coating. Technol. 153 (2002) 166–172, https://doi.org/10.1016/S0257-8972(01)01665-6.

[21] A. Seynstahl, S. Krauß, E. Bitzek, B. Meyer, B. Merle, S. Tremmel, Microstructure, mechanical properties and tribological behavior of magnetron-sputtered MoS2 solid lubricant coatings deposited under industrial conditions, Coatings 11 (2021) 455, https://doi.org/10.3390/coatings11040455.

[22] J. Moser, F. Lévy, Crystal reorientation and wear mechanisms in MoS2 lubricating thin films investigated by TEM, J. Mater. Res. 8 (1993) 206–213, https://doi.org/10.1017/S0884291400120539.

[23] F. Lévy, J. Moser, High-resolution cross-sectional studies and properties of molybdenite coatings, Surf. Coating. Technol. (1994) 68–69, https://doi.org/10.1016/0257-8972(94)90198-8, 433–438.

[24] J.M. Martin, H. Pascal, C. Donnet, T. Le Mogne, J.L. Loubet, T. Epicier, Superlubricity of MoS2: crystal orientation mechanisms, Surf. Coating. Technol. 68–69 (1994) 427–432, https://doi.org/10.1016/0257-8972(94)90197-X.

[25] E. Mishina, N. Sherstyuk, S. Lavrov, A. Sigov, A. Mitioglu, S. Anghel, L. Kulyuk, Observation of two polytypes of MoS2 ultrathin layers studied by second harmonic generation microscopy and photoluminescence, Appl. Phys. Lett. 106 (2015) 131901, https://doi.org/10.1063/1.4907972.

[26] R.W.G. Wyckoff, Crystal Structures, second ed., vol. 1, 1963, pp. 280–281.

[27] J.F. Archard, Contact and rubbing of flat surfaces, J. Appl. Phys. 24 (1953) 981–988, https://doi.org/10.1063/1.1721448.

[28] W. Tillmann, A. Wittig, D. Stangier, H. Moldenhauer, C.-A. Thomann, J. Debus, D. Aurich, A. Bruemmer, Influence of the bias-voltage, the argon pressure and the heating power on the structure and the tribological properties of HiPIMS sputtered MoSx films, Surf. Coating. Technol. 385 (2020) 125358, https://doi.org/10.1016/j.surfcoat.2020.125358.

[29] J. Wang, W. Lauwerens, E. Wieers, L.M. Stals, J. He, J.P. Celis, Structure and tribological properties of MoSx coatings prepared by bipolar DC magnetron sputtering, Surf. Coating. Technol. 139 (2001) 143–152, https://doi.org/10.1016/S0257-8972(01)00988-4.

[30] V. Buck, A neglected parameter (water contamination) in sputtering of MoS2 films, Thin Solid Films 139 (1986) 157–168, https://doi.org/10.1016/0040-6090(86)90334-2.

[31] J. Moser, F. Lévy, F. Bussy, Composition and growth mode of MoSx sputtered films, J. Vac. Sci. Technol. Vac. Surf. Films. 12 (1994) 494–500, https://doi.org/10.1116/1.579157.

[32] D.-Y. Wang, C.-L. Chang, W.-Y. Ho, Microstructure analysis of MoS2 deposited on diamond-like carbon films for wear improvement, Surf. Coating. Technol. 111 (1999) 123–127, https://doi.org/10.1016/S0257-8972(98)00712-9.

[33] T. Spalvins, Lubrication with sputtered MoS2 films: principles, operation, and limitations, J. Mater. Eng. Perform. 1 (1992) 347–351, https://doi.org/10.1007/BF02652388.

[34] B. Vierneusel, S. Tremmel, S. Wartzack, Monte Carlo simulation of the MoS2 sputtering process and the influence of the normalized momentum on residual stresses, J. Vac. Sci. Technol. Vac. Surf. Films. 33 (2015) 061501, https://doi.org/10.1116/1.4926383.

[35] K. Müller, Stress and microstructure of sputter-deposited thin films: molecular dynamics investigations, J. Appl. Phys. 62 (1987) 1796–1799, https://doi.org/10.1063/1.339559.

[36] P. Serles, H. Sun, G. Colas, J. Tam, E. Nicholson, G. Wang, J. Howe, A. Saulot, C. V. Singh, T. Filleter, Structure-dependent wear and shear mechanics of nanostructured MoS2 coatings, Adv. Mater. Interfac. 7 (2020) 1901870, https://doi.org/10.1002/admi.201901870.

[37] K. Hou, S. Yang, X. Liu, J. Wang, The self-ordered lamellar texture of MoS2 transfer film formed in complex lubrication, Adv. Mater. Interfac. 5 (2018) 1701682, https://doi.org/10.1002/admi.201701682.

[38] C. Donnet, J.M. Martin, T. Le Mogne, M. Belin, Super-low friction of MoS2 coatings in various environments, Tribol. Int. 29 (1996) 123–128, https://doi.org/10.1016/0301-679X(95)00094-K.

[39] G. Theiler, T. Gradt, W. Österle, A. Brückner, V. Weihnacht, Friction and endurance of MoS2/ta-C coatings produced by Laser Arc deposition, Wear 297 (2013) 791–801, https://doi.org/10.1016/j.wear.2012.10.007.

[40] Y. Shi, J. Pu, L. Wang, Structural phase transformation in amorphous molybdenum disulfide during friction, J. Phys. Chem. C 125 (2021) 836–844, https://doi.org/10.1021/acs.jpcc.0c06315.

[41] K.L. Johnson, Contact Mechanics, first ed., Cambridge University Press, 1985 https://doi.org/10.1017/CBO9781139171731.

[42] T. Gradt, T. Schneider, Tribological performance of MoS$_2$ coatings in various environments, Lubricants 4 (2016) 32, https://doi.org/10.3390/lubricants4030032.

[43] S.S. Jo, A. Singh, L. Yang, S.C. Tiwari, S. Hong, A. Krishnamoorthy, M.G. Sales, S. M. Oliver, J. Fox, R.L. Cavalero, D.W. Snyder, P.M. Vora, S.J. McDonnell, P. Vashishta, R.K. Kalia, A. Nakano, R. Jaramillo, Growth kinetics and atomistic mechanisms of native oxidation of ZrSx Se2–x and MoS2 crystals, Nano Lett 20 (2020) 8592–8599, https://doi.org/10.1021/acs.nanolett.0c03263.

[44] J.F. Curry, M.A. Wilson, H.S. Luftman, N.C. Strandwitz, N. Argibay, M. Chandross, M.A. Sidebottom, B.A. Krick, Impact of microstructure on MoS2 oxidation and friction, ACS Appl. Mater. Interfaces 9 (2017) 28019–28026, https://doi.org/10.1021/acsami.7b06917.